\documentclass[iop,apjl]{emulateapj}

\usepackage{graphicx}  
\usepackage{dcolumn}   
\usepackage{bm}        
\usepackage{amsfonts,amsmath,amssymb,mathrsfs}
\usepackage{color}
\usepackage{epsfig}
\usepackage{natbib,times}
\usepackage{url}
\usepackage{times}

\pdfoutput=1

\def\VEC#1{\mbox{\boldmath $#1$}}

\makeatletter
\newcommand{\vast}{\bBigg@{4}}
\newcommand{\Vast}{\bBigg@{5}}
\makeatother

\DeclareMathAlphabet{\mathpzc}{OT1}{pzc}{m}{it}

\shorttitle{Magnetic Reconnection in the Ergosphere of a non-Kerr Black Hole}
\shortauthors{Wenshuai Liu}

\begin{document}

\title{Energy Extraction via Magnetic Reconnection in the Ergosphere of a rotating non-Kerr Black Hole}

\author{Wenshuai {Liu}\altaffilmark{1}$^{*}$}

\altaffiltext{1}{Department of Astronomy, School of Physics, Huazhong University of Science and Technology, Wuhan 430074, China}
\altaffiltext{*}{E-mail: 674602871@qq.com}

\begin{abstract}
Magnetic reconnection process in the ergosphere is investigated for a relativistic plasma around
a rotating non-Kerr black hole. For a rotating non-Kerr black hole immersed in magnetic field generated by an externally material, anti-parallel magnetic field line could form in the ergosphere due to the frame dragging. Therefore, magnetic reconnection could occur in the ergosphere. Such magnetic reconnection may generate negative energy at infinity by redistributing the angular momentum during the process. The results show that, with the effect of the deformed parameter, extraction of energy from a rotating non-Kerr black hole by magnetic reconnection could be enhanced in the presence of a positive deformed parameter.
\end{abstract}

\keywords{black hole physics --- plasma astrophysics --- magnetic fields}

\section{Introduction}

According to the no-hair theorem \citep{60,61,62,63,64}, it shows that a rotating black hole with no electric charge is fully described by the Kerr metric with the mass of the black hole $M$ and the dimensionless spin parameter $a$ in asymptotically flat and matter-free spacetime. However, general relativity may break down in the case of strong gravity, meaning that Kerr metric might not be the unique spacetime accounting for rotating black holes \citep{65,66,67}. With this motivation, a deformed Kerr-like metric is proposed by Johannsen and Psaltis that is suitable to describe a rotating non-Kerr black hole in the regime of strong gravity \citep{66}. Compared with Kerr metric, non-Kerr metric contains a deformed parameter $\epsilon$ which represents deviations from Kerr metric.

Rotating black holes store rotational energy that can be extracted \citep{68}. Extraction of energy from a rotating black hole could produce significant energy outflows which are believed to account for some kinds of high energy astrophysical processes, such as relativistic jets powered by black holes hosted in active galactic nuclei (AGNs), gamma-ray bursts (GRBs) and microquasars ($\mu$QSOs) \citep{69}. Various processes of extracting energy from rotating black holes have been developed. Mechanisms accounting for energy extraction of rotating black holes include Penrose process, superradiant scattering, Blandford-Znajek mechanism, magnetohydrodynamic Penrose process and modified Hawking process \citep{70,71,72,73,74,75}. Especially, these different kinds of mechanisms of energy extraction are based on the idea of Penrose process that it redistributes energy and angular momentum of which the negative part swallowed by the black hole will reduce the black hole mass. Recently, magnetic reconnection in the ergosphere of a Kerr black hole \citep{76,77} shows to be an efficient way of extracting energy from a Kerr black hole by producing a pair of outflows with velocity of equal values but opposite directions through redistributing angular momentum of plasma in a localized diffusion region with one part of negative energy at infinity and the other part of the energy which is larger than its rest mass and thermal energies, acting as the same way as that in Penrose process.

Energy extraction of rotating black holes listed above is proposed in the regime of Kerr metric. Energy extraction by Penrose process in the ergosphere of a rotating non-Kerr back hole is investigated by \cite{78} where they showed that the deformed parameter $\epsilon$ in the non-Kerr metric could play a significant role on the maximum efficiency of energy extraction due to Penrose process and could cause the efficiency to exceed to $60\%$ compared with $\sim 20.7\%$ in the ergosphere of Kerr metric. In this work, we investigate in detail energy extraction via magnetic reconnection in the equatorial plane in the ergosphere of a rotating non-Kerr black hole and how the deformed parameter $\epsilon$ affects the efficiency of the energy extraction due to magnetic reconnection.

This work is organized as follows: we briefly describe the non-Kerr metric proposed by Johannsen and Psaltis and the physical background with which to study energy extraction due to magnetic reconnection in the ergosphere of a rotating non-Kerr black hole in Section 2. We investigate the efficiency of the energy extraction via magnetic reconnection in Section 3. The discussion is in Section 4 and a summary is given in Section 5. Throughout this work, we set gravitational constant $G=1$, mass of the non-Kerr black hole $M=1$ and the speed of light $c=1$.

\section{The physical background of magnetic reconnection in the non-Kerr spacetime}
\begin{figure}[t!]
\begin{center}
\includegraphics[clip,width=0.5\textwidth]{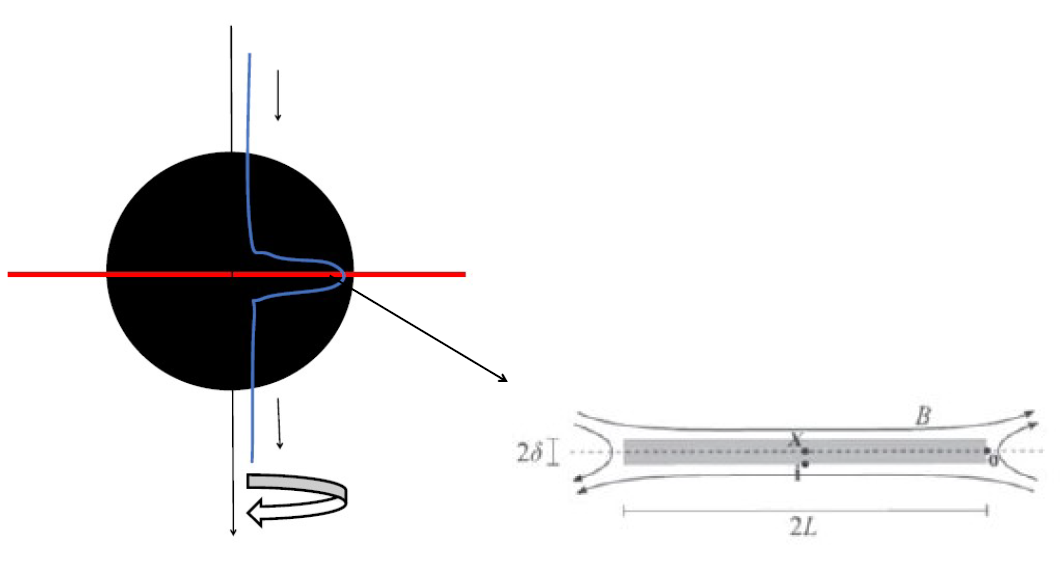}
\end{center}
\caption{ Magnetic configuration of the magnetic reconnection in the black hole ergosphere. The red line represents the equatorial plane of the non-Kerr black hole. The blue curve which is behind the black hole, shows the magnetic field line affected by the frame dragging effect.}
\label{fig:figure11}
\end{figure}

Same as extraction of energy from a Kerr black hole in the ergosphere, energy extraction of a rotating non-Kerr black hole through the negative-energy particles swallowed by the black hole via magnetic reconnection in the ergosphere needs the condition that it occurs in the ergosphere with the fact that negative-energy orbits only exist in such region. Configuration of the antiparallel magnetic field lines with which magnetic reconnection could take place is a natural result of frame dragging effect as shown in Figure.~\ref{fig:figure11}. Before investigating magnetic reconnection in the ergosphere of a rotating non-Kerr black hole, we first describe the physical background with which to study energy extraction via magnetic reconnection in the ergosphere.

The deformed Kerr-like metric around a rotating non-Kerr black hole in Boyer-Lindquist coordinates ($t,r,\theta,\phi$) can be expressed as \citep{66}
\begin{eqnarray}
ds^2=g_{tt}dt^2+g_{rr}dr^2+g_{\theta\theta}d\theta^2+g_{\phi\phi}
d\phi^2+2g_{t\phi}dtd\phi, \label{metric0}
\end{eqnarray}
with
\begin{eqnarray}
g_{tt}&=&-\bigg(1-\frac{2Mr}{\rho^2}\bigg)(1+h),\;\;\;\;\;
g_{t\phi}=-\frac{2aMr\sin^2\theta}{\rho^2}(1+h),\nonumber\\
g_{rr}&=&\frac{\rho^2(1+h)}{\Delta+a^2h\sin^2\theta},\;\;\;\;\;\;\;\;\;\;\;\;\;\;\;
g_{\theta\theta}=\rho^2,\nonumber\\
g_{\phi\phi}
&=&\sin^2\theta\bigg[\rho^2
+\frac{a^2(\rho^2+2Mr)\sin^2\theta}{\rho^2}(1+h)\bigg],
\end{eqnarray}
where
\begin{eqnarray}
\rho^2&=&r^2+a^2\cos^2\theta,\;\;\;\;\;\;\;\;\;\;
\Delta=r^2-2Mr+a^2,\nonumber\\
h&=&\frac{\epsilon M^3
r}{\rho^4}.
\end{eqnarray}
and $r,\theta,\phi$ represent the radial distance, the polar angle and the azimuthal angle, respectively.

The constant $\epsilon$ represents the deformation parameter. $\epsilon$ with positive or negative value corresponds to the cases in which the compact object is more prolate or oblate than the Kerr black hole whose $\epsilon$ is equal to zero.

Energy extraction can happen in the ergosphere region of which is bounded by the event horizon $r_H$ and the outer infinite redshift surface $r_\infty$. The horizon of the black hole $r_H$ is given by the maximum root of the following equation \citep{66}

\begin{equation}
\Delta+a^2h\sin^2\theta=0
\end{equation}
and the outer infinite redshift surface is described by \citep{78}

\begin{equation}
r_\infty=M+(M^2-a^2\cos^2\theta)^\frac{1}{2}
\end{equation}
which is the root of the following equation

\begin{equation}
1-\frac{2Mr}{\rho^2}=0
\end{equation}

Due to the presence of $\epsilon$, the non-Kerr black hole could have an ergosphere only when \citep{78}

\begin{equation}
-4(M+\sqrt{M^2-a^2\cos^2\theta})\leq\epsilon\leq
\frac{\Delta\rho^4}{M^3a^2r\sin^2\theta}\bigg|_{r=r_{tp}}.
\end{equation}
where $r_{tp}$ is the maximum positive root of the following equation \citep{78}

\begin{equation}
10 r^4-16 M r^3+a^2(7+\cos(2\theta))r^2-a^4(1+\cos(2\theta))=0.
\end{equation}

In what follows, we study energy extraction using 3+1 formalism. 3+1 formalism with the advantage that it can isolate the effects of rotation of the spacetime, is suitable to study plasmas surrounding rotating compact objects. In order to study the process of magnetic reconnection in the ergosphere of the rotating non-Kerr black hole analytically, we adopt a locally nonrotating frame called zero angular momentum observer (ZAMO) frame \citep{79} for convenience. Then we have the line element

\begin{equation}
ds^2= - d \hat{t}^2 + \sum _i (d \hat{x}^i)^2
= \eta_{\mu\nu} d\hat{x}^\mu d\hat{x}^\nu
\end{equation}
where $d\hat{t} = \alpha dt$, $d\hat{x}^i = h_i dx^i - \alpha \beta ^i dt$, $\alpha=\sqrt{-g_{tt}+\frac{g_{t\phi}^2}{g_{\phi \phi}}}$ and $\beta^\phi=-\frac{g_{t\phi}}{\alpha\sqrt{g_{\phi \phi}}}$.

Obviously, the spacetime is identical to Minkowski spacetime locally in the ZAMO frame. For a contravariant vector $a^\mu$ in the Boyer-Lindquist coordinates, $\hat{a}^\mu$ in the ZAMO frame is
\begin{equation}
\hat{a}^0 = \alpha a^0, \verb!   !
\hat{a}^i = h_i a^i - \alpha \beta ^i a^0
\end{equation}

and the covariant vector $\hat{a}_\mu$ is
\begin{equation}
\hat{a}_0 = \frac{1}{\alpha} a_0 + \sum _i \frac{\beta ^i}{h_i} a_i, \verb!   !
\hat{a}_i = \frac{1}{h_i} a_i
\end{equation}

Based on these transformations, we get the energy-at-infinity density $e^\infty = - \alpha g_{\nu 0} T^{\nu 0}$ with variables with hat observed in the ZAMO frame as follow \citep{76}
\begin{equation}
e^\infty = \alpha \hat{e} + \sum_i \omega_i h_i \hat{P}^i =
\alpha \hat{e} + \alpha\beta^\phi \hat{P}^\phi
\label{3}
\end{equation}
where $\hat{e}$ and $\hat{P}^\phi$ are the total energy density and the azimuthal component of the momentum density, respectively. They are given as \citep{76}

\begin{eqnarray}
\hat{e} = {\mathfrak h} \hat{\gamma}^2 -p + \frac{\hat{B}^2 + \hat{E}^2}{2}\\
\hat{P}^\phi ={\mathfrak h}\hat{\gamma}^2 \hat{v}^\phi
+ ( \hat{\VEC{E}} \times \hat{\VEC{B}})^\phi
\end{eqnarray}
where $\mathfrak h$ is the enthalpy density.

In order to clearly show the energy-at-infinity density with hydrodynamic and electromagnetic part, equation (\ref{3}) could be separated as $e^\infty=e^\infty_{\rm hyd}+e^\infty_{\rm EM}$ where \citep{76}
\begin{eqnarray}
e^\infty_{\rm hyd} = \alpha \hat{e}_{\rm hyd} +\alpha \beta^\phi {\mathfrak h}\hat{\gamma}^2\hat{v}^\phi \label{4} \\
e^\infty_{\rm EM}  = \alpha \hat{e}_{\rm EM}+\alpha \beta^\phi( \hat{\VEC{E}} \times \hat{\VEC{B}})^\phi \label{5}
\end{eqnarray}
where $\hat{e}_{\rm hyd}={\mathfrak h} \hat{\gamma}^2 -p$ and $\hat{e}_{\rm EM}=(\hat{B}^2 + \hat{E}^2)/2$ represent the hydrodynamic and electromagnetic energy densities observed
in the ZAMO frame, respectively.

Configuration of anti-parallel magnetic field lines near the equatorial plane in the ergosphere resulting from the frame dragging is considered here. For simplicity, we only investigate the hydrodynamic energy density at infinity with the assumption that magnetic reconnection converts most of the magnetic energy into plasma particle energy \citep{77}. Thus, energy at infinity is mainly dominated by hydrodynamic energy at infinity. Therefore, equation (\ref{4}) changes to \citep{76}
\begin{equation}
e^\infty_{\rm hyd}=\alpha\left[{\mathfrak h}(\hat{\gamma}+\beta^\phi\hat{\gamma}\hat{v}^\phi)-\frac{p}{\hat{\gamma}} \right] \label{66}
\end{equation}
with the condition that the plasma is incompressible and adiabatic.

We assume the fluid element in the plasma co-rotates around the rotating non-Kerr black hole in the equatorial plane with Keplerian velocity which can be expressed as following in Boyer-Lindquist coordinates \citep{67}
\begin{equation}
\Omega=\frac{d\phi}{dt}=\frac{-g_{t \phi,r}+\sqrt{g_{t \phi, r}^2-g_{tt,r}g_{\phi \phi,r}}}{g_{\phi \phi, r}}
\end{equation}

The above angular velocity could be written as the Keplerian velocity in the $\phi$ direction observed in the ZAMO frame based on the transformation between a contravariant vector in the Boyer-Lindquist coordinates and that in the ZAMO frame
\begin{equation}
\hat v_K=\frac{\Omega \sqrt{g_{\phi \phi}}}{\alpha}-\beta^\phi
\end{equation}

Circular orbits in the equatorial plane can exist from $r \rightarrow \infty$ to the circular photon orbit which occurs at the radius at which \citep{66}

\begin{equation}
\frac{E}{\mu} = \frac{1}{r^6}\sqrt{ \frac{P_1 + P_2}{P_3} }\rightarrow\infty \label{6060}
\end{equation}
and
\begin{eqnarray}
\frac{L_z}{\mu} = &&  \frac{1}{r^4 P_6 \sqrt{P_3}} \big[ \sqrt{M(r^3+\epsilon_3 M^3) P_5 } \nonumber \\
&& - 6a M (r^3 + \epsilon_3 M^3) \sqrt{P_1 + P_2} \big]\rightarrow\infty \label{7070}
\end{eqnarray}
where $\mu$ is the rest mass of photon.

From equation (\ref{6060}) and (\ref{7070}), we could get the radius where the circular photon orbit $r_{ph}$ exists by solving $P_3=0$. However, circular orbits for a particle are stable only when $r>r_{isco}$ where $r_{isco}$ is the innermost stable circular orbit for nonspinning test particles, meaning that not all circular orbits are stable for $r>r_{ph}$. For simplicity, we don't consider the stability of the circular orbit in this work.

\begin{figure*}
   \begin{center}
     \begin{tabular}{cc}
     \includegraphics[width=0.33\textwidth]{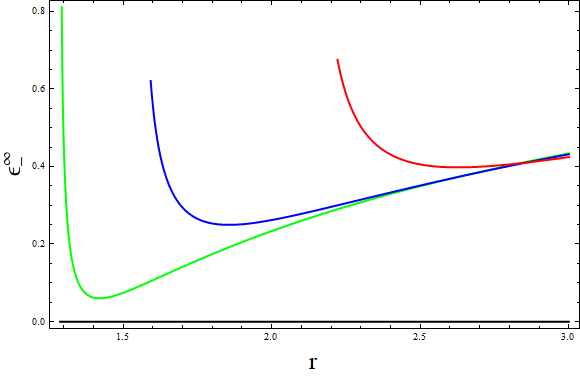}
     \includegraphics[width=0.33\textwidth]{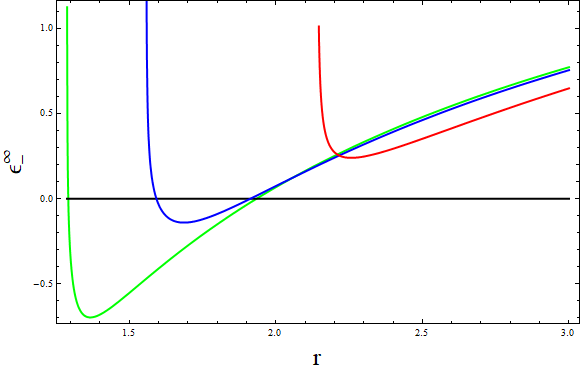}
     \includegraphics[width=0.33\textwidth]{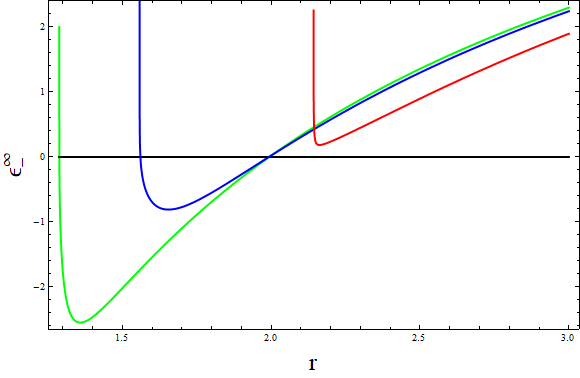}\\
     \includegraphics[width=0.33\textwidth]{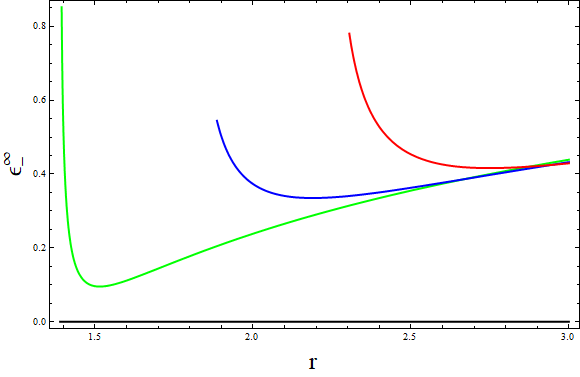}
     \includegraphics[width=0.33\textwidth]{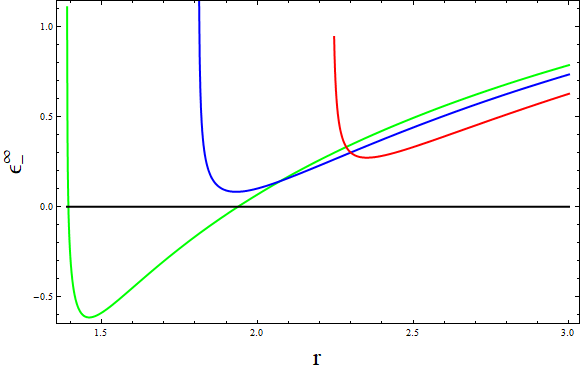}
     \includegraphics[width=0.33\textwidth]{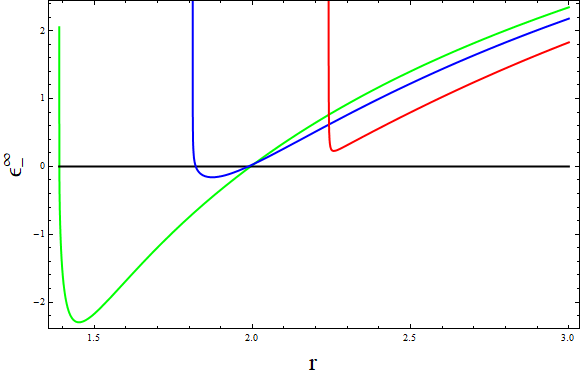}\\
     \includegraphics[width=0.33\textwidth]{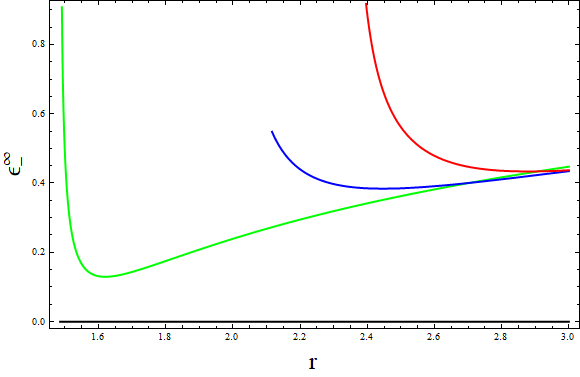}
     \includegraphics[width=0.33\textwidth]{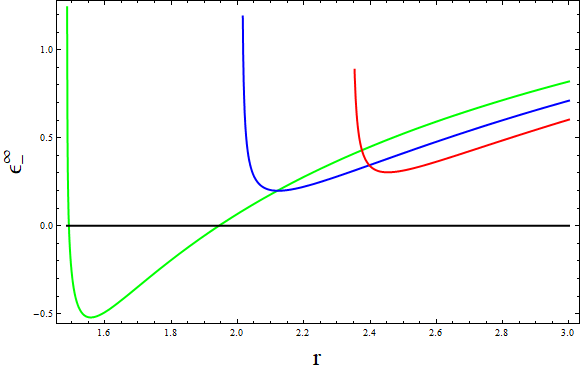}
     \includegraphics[width=0.33\textwidth]{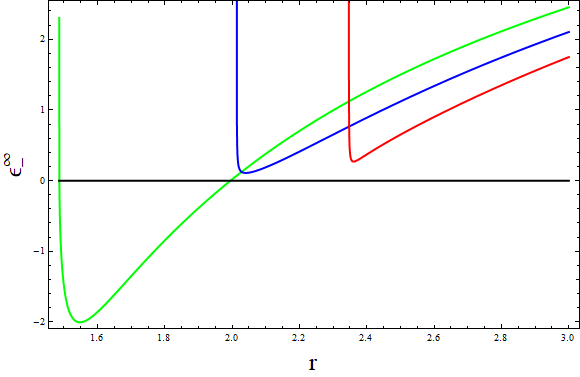}
     \end{tabular}
   \end{center}

    \caption{$\epsilon^\infty_{-}$ as a function of $r$ with black hole spin $a=0.9$ (up panel) $a=0.8$ (middle panel) and $a=0.7$ (bottom panel) and plasma magnetization $\sigma_0=1,10,100$ from left to right, respectively. Red and blue curves represent the results with $\epsilon=-1.8$ and $\epsilon=0$, respectively. Green curves in the up panel, middle panel and bottom panel represent the results with $\epsilon=0.3$, $\epsilon=0.9$ and $\epsilon=1.9$, respectively.}
    \label{fig:figure1}
\end{figure*}

In order to obtain the outflow velocity via the magnetic reconnection in small spatial scale, we introduce the local rest frame  $(t',x^{1'},x^{2'},x^{3'})$ co-moving with the bulk plasma which rotates around the non-Kerr black hole in the circular orbit in the equatorial plane. The local rest frame is set with the condition that the direction of $x^{1'}$ is parallel to the direction $r$ and the direction of $x^{3'}$ is parallel to the direction $\phi$. The outflow velocity $v^{\phi '}$ and the outflow Lorentz factor $\gamma '$ observed in the local rest frame could be changed to $\hat v^\phi$ and $\hat \gamma$ observed in the ZAMO frame through the following transformation \citep{80}

\begin{equation}
\hat a^\mu=e_{\nu '}^{\mu} a^{\nu '}
\end{equation}

where

\begin{eqnarray}
&&\left(
	\begin{array}{cccc}
	e^{\hat{t}}_{~t'} & e^{\hat{t}}_{~r'}
		& e^{\hat{t}}_{~\theta'} & e^{\hat{t}}_{~\phi'} \\
	e^{\hat{r}}_{~t'} & e^{\hat{r}}_{~r'}
		& e^{\hat{r}}_{~\theta'} & e^{\hat{r}}_{~\phi'} \\
	e^{\hat{\theta}}_{~t'} & e^{\hat{\theta}}_{~r'}
		& e^{\hat{\theta}}_{~\theta'} & e^{\hat{\theta}}_{~\phi'} \\
	e^{\hat{\phi}}_{~t'} & e^{\hat{\phi}}_{~r'}
		& e^{\hat{\phi}}_{~\theta'} & e^{\hat{\phi}}_{~\phi'} \\
	\end{array}
\right)
=\nonumber\\&&\left(
	\begin{array}{cccc}
	\hat{\gamma}
		& \hat{\gamma}\hat{v}_r
		& 0
		& \hat{\gamma}\hat{v}_\phi \\
	\hat{\gamma}\hat{v}_r
		& \displaystyle 1+\frac{\hat{\gamma}^2\hat{v}_r^2}{1+\hat{\gamma}}
		& 0
		& \displaystyle
		\frac{\hat{\gamma}^2\hat{v}_r\hat{v}_\phi}{1+\hat{\gamma}} \\
	0
		& 0
		& 1
		& 0 \\
	\hat{\gamma}\hat{v}_\phi
		& \displaystyle \frac{\hat{\gamma}^2\hat{v}_r
			\hat{v}_\phi}{1+\hat{\gamma}}
		& 0
		& \displaystyle
			1+\frac{\hat{\gamma}^2 \hat{v}_\phi^2}{1+\hat{\gamma}}
	\end{array}
\right)\label{77}
\end{eqnarray}
$\hat v_r$ and $\hat v_\phi$ are the local rest frame's velocity in the radial and $\phi$ direction observed in ZAMO frame, respectively. In this work where the local rest frame co-moves with the bulk plasma rotating around the non-Kerr black hole in the circular orbit in the equatorial plane, it shows that $\hat v_r =0$ and $\hat v_\phi = \hat v_K$.

We adopt the configuration of reconnection layer in the azimuthal direction with the radially oriented current density, then we get the outflow velocity observed in the local rest frame as \citep{77}

\begin{equation}
v_{out}=(\frac{\sigma_0}{1+\sigma_0})^\frac{1}{2} \label{88}
\end{equation}
where $\sigma_0=B_0^2/{\mathfrak h}_0$ is the plasma magnetization upstream of the reconnection layer, $B_0$ is the asymptotic macro-scale magnetic field and ${\mathfrak h}_0$ is the enthalpy density \citep{77}.

With equation (\ref{88}) and (\ref{77}), we get the outflow velocity observed in the ZAMO frame

\begin{equation}
v_\pm^\phi=\frac{\hat v_K \pm v_{out}}{1 \pm \hat v_K v_{out}} \label{99}
\end{equation}
where $\pm$ represent the outflow velocity with corotating (+) and counterrotating (-) direction relative to the rotation of the non-Kerr black hole.

\begin{figure*}
   \begin{center}
     \begin{tabular}{cc}
     \includegraphics[width=0.5\textwidth]{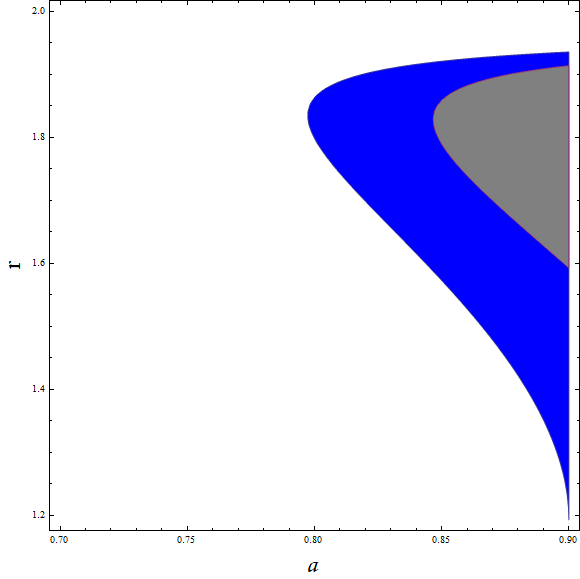}
     \includegraphics[width=0.5\textwidth]{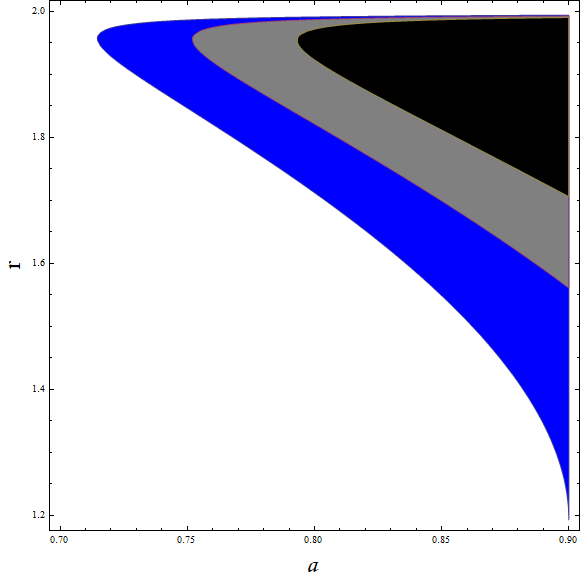}
     \end{tabular}
   \end{center}

    \caption{Regions of the phase-space $\{a,r\}$ where $\epsilon^\infty_{-} < 0$ with plasma magnetization $\sigma_0=10$ (left panel) and $\sigma_0=100$ (right panel). The blue, the gray and the black regions represent the results with $\epsilon=0.3202$, $\epsilon=0$ and $\epsilon=-0.3$ respectively.}
    \label{fig:figure2}
\end{figure*}

\section{Extraction of energy from a non-Kerr black hole via magnetic reconnection}

With the above equation (\ref{99}) and equation (\ref{66}), we get the energy-at-infinity density of the reconnection outflows observed in the ZAMO frame as

\begin{eqnarray}
e^\infty_{{\rm hyd},\pm} \!=\!&& \alpha \hat\gamma_K \Bigg[ \left(1 \!+\! \hat v_K  \beta^\phi  \right)  \gamma_{out} {\mathfrak h}
  \pm   \left(\hat v_K \!+\!  \beta^\phi \right)  \gamma_{out}  v_{out} {\mathfrak h}  \nonumber \\
 &&-\frac{p}{\left(1 \!\pm\!  \, \hat v_K   v_{out} \right) \gamma_{out}  \hat\gamma_K^2}  \Bigg] \, ,
\end{eqnarray}
where $ \gamma_{out}=(1- v_{out}^2)^{-\frac{1}{2}}$ and $\pm$ represent energy-at-infinity density of the accelerated (+) outflow and decelerated (-) outflow relative to the rotation of the black hole.

Then the hydrodynamic energy at infinity per enthalpy is \citep{77}

\begin{eqnarray}
\epsilon^\infty_{\pm} =\frac{e^\infty_{{\rm hyd},\pm}}{{\mathfrak h}}\!=\!&& \alpha \hat\gamma_K \Bigg[ \left(1 \!+\! \hat v_K  \beta^\phi  \right) \gamma_{out}
  \pm   \left(\hat v_K \!+\!  \beta^\phi \right) \gamma_{out}  v_{out}   \nonumber \\
 &&-\frac{1}{4\left(1 \!\pm\!  \, \hat v_K   v_{out} \right)  \gamma_{out}  \hat\gamma_K^2}  \Bigg] \, ,
 \label{10}
\end{eqnarray}
where we set $\Gamma=\frac{4}{3}$.

With equation (\ref{10}), we could get the energy-at-infinity per enthalpy of the two outflows observed in the ZAMO frame after magnetic reconnection in the ergosphere of a rotating non-Kerr black hole. After magnetic reconnection, the interesting phenomenon is that whether the energy-at-infinity of the decelerated outflow observed in the ZAMO frame could be negative or not. Since energy extraction through magnetic reconnection in the ergosphere of a Kerr black hole is studied in detail, here, we focus on how the deformed parameter in the non-Kerr metric affects the energy of the decelerated outflow after magnetic reconnection in the ergosphere and the efficiency of the energy extraction.

In Figure.~\ref{fig:figure1} the energy-at-infinity of the decelerated plasma observed in the ZAMO frame is shown for different values of the deformed parameter $\epsilon$, black hole spin $a$ and plasma magnetization $\sigma_0 $. $\epsilon$ with posive/negative value is responsible for shifting $r_{ph}$ towards/outwards the central black hole, meaning that the minimum radius of circular orbit of plasma could decrease/increase when $\epsilon >0$/$\epsilon <0$. As seen from Figure.~\ref{fig:figure1}, the radial region where energy extraction via magnetic reconnection could occur becomes large as $\epsilon$ increases for the same $\sigma_0 $.

\begin{figure*}
   \begin{center}
     \begin{tabular}{cc}
     \includegraphics[width=0.33\textwidth]{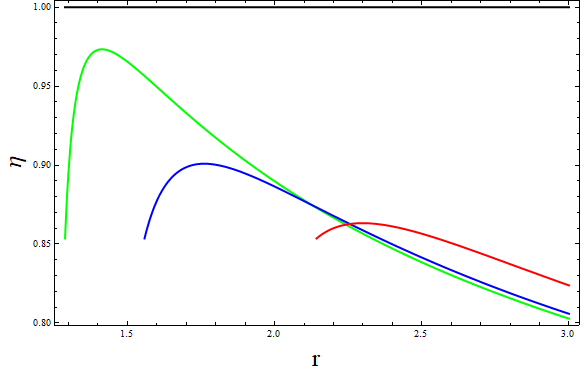}
     \includegraphics[width=0.33\textwidth]{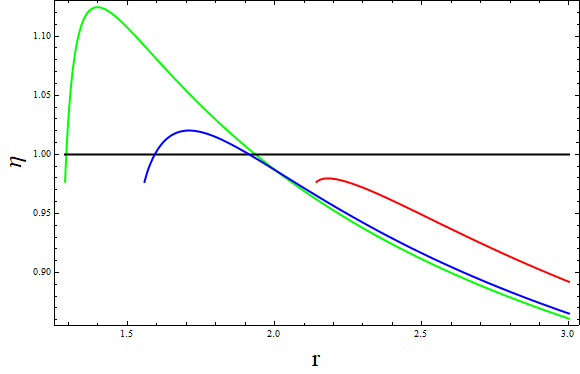}
     \includegraphics[width=0.33\textwidth]{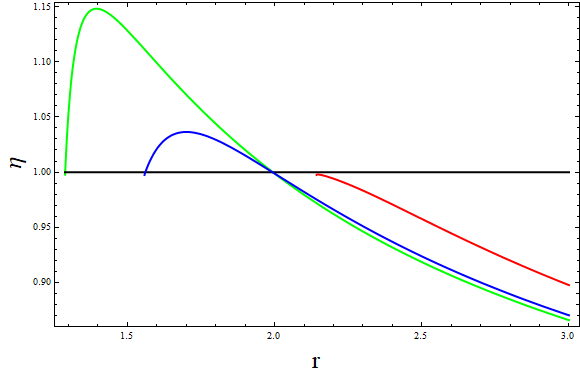}\\
     \includegraphics[width=0.33\textwidth]{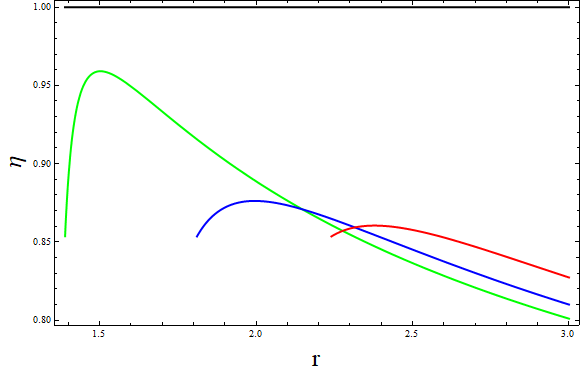}
     \includegraphics[width=0.33\textwidth]{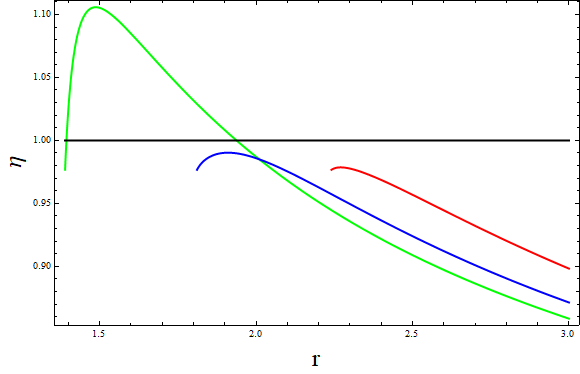}
     \includegraphics[width=0.33\textwidth]{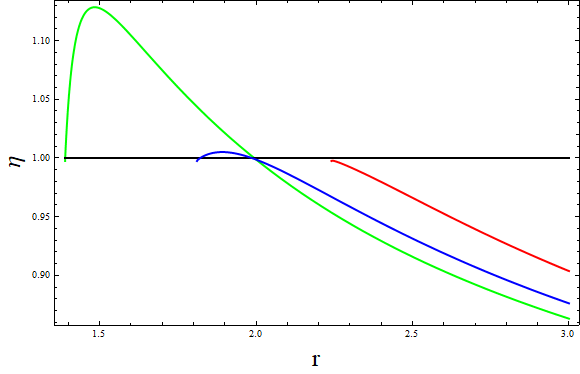}\\
     \includegraphics[width=0.33\textwidth]{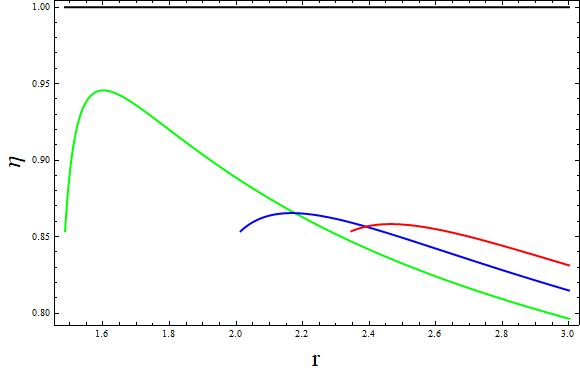}
     \includegraphics[width=0.33\textwidth]{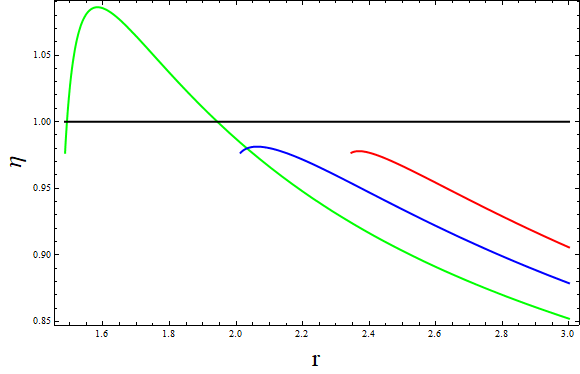}
     \includegraphics[width=0.33\textwidth]{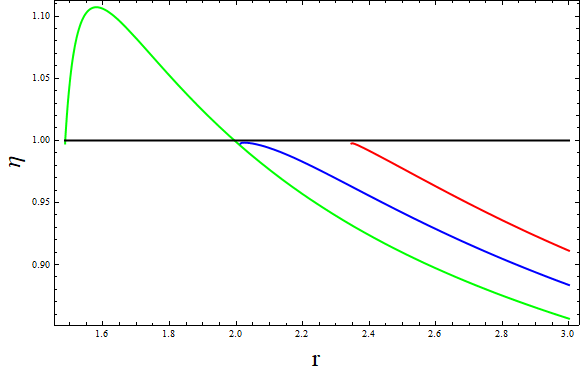}
     \end{tabular}
   \end{center}

    \caption{$\eta$ as a function of $r$ with black hole spin $a=0.9$ (up panel) $a=0.8$ (middle panel) and $a=0.7$ (bottom panel) and plasma magnetization $\sigma_0=1,10,100$ from left to right, respectively. Red and blue curves represent the results with $\epsilon=-1.8$ and $\epsilon=0$, respectively. Green curves in the up panel, middle panel and bottom panel represent the results with $\epsilon=0.3$, $\epsilon=0.9$ and $\epsilon=1.9$, respectively.}
    \label{fig:figure3}
\end{figure*}

In Figure.~\ref{fig:figure2}, the region of the phase-space $\{a,r\}$ where rotational energy of a rotating non-Kerr black hole could be extracted by magnetic reconnection decreases as $\epsilon$ decreases for the same $\sigma_0 $.

In order to illustrate the efficiency of energy extraction due to magnetic reconnection, we define \citep{77}
\begin{equation}
\eta=\frac{\epsilon^\infty_{+}}{\epsilon^\infty_{+}+\epsilon^\infty_{-}}
\end{equation}
as the efficiency of extraction of energy via magnetic reconnection. Energy extraction happens when $\eta >1$.

Figure.~\ref{fig:figure3} presents the efficiency $\eta$ as a function of the radius with different $\epsilon$, $a$ and $\sigma_0 $. It shows that the efficiency increases as $\epsilon$ increases.

In real astrophysical systems hosting black holes, it is expected that the plasma magnetization in the ergosphere could be $\sigma_0\geq1$. Especially, in the case of supermassive black hole (SMBH) in active galactic nuclei, the magnetization of plasma around such SMBH could reach $\sigma_0\sim 10^4$ or even larger \citep{118,81,82,83}. In the condition of stellar mass black hole governing gamma-ray bursts, the plasma magnetization in the ergosphere could be $\sigma_0\sim 1$ or larger \citep{84,75,85,86}. With plasma magnetization of high value around a rotating non-Kerr black hole, efficient energy extraction of non-Kerr black holes via magnetic reconnection in the ergosphere could occur. It shows that the spin of the supermassive black hole in M87 is about 0.9 \citep{102}, thus, it is expected that efficient extraction of energy via magnetic reconnection in the ergosphere of M87* could take place if the deformed parameter is positive compared with that with Kerr metric. Many methods have been proposed to test the no-hair theorem in the strong field regime \citep{103,104,105,106,107,108,109,110,111,112,113,114,115,116,117}. There is still a lack of accurate constrains on the deformed parameter of a given black hole based on observation at present. From the theoretical point of view, magnetic reconnection in the ergosphere of a rotating non-Kerr black hole with large spin $a$, large plasma magnetization $\sigma_0$ and large deformed parameter $\epsilon$ with positive value leads to extraction of energy with high efficiency.

\section{Discussion}
The supermassive black hole in M87 shows bright TeV flares \citep{91,92,93,94}. It is conjectured that dissipation of magnetic energy resulting from magnetic reconnection is responsible for triggering flares powered by the energetic electrons \citep{95,96,97,98,99,100,101}. Results from this work show that the deformed parameter could play a significant role on the outflow energy through magnetic reconnection in contrast with the Kerr spacetime. Plasmoids resulting from outflows produced by magnetic reconnection merge with other plasmoids, forming hot spots. Such hot spots filled with electrons energized by magnetic reconnection can power high-energy flares. Thus, features of the flares may be different between the non-Kerr spacetime and the Kerr spacetime with the same black hole spin and plasma magnetization, meaning that the features may be used to distinguish the non-Kerr black hole from Kerr black hole. Such features are based on magnetic reconnection simulations with general relativistic magnetohydrodynamics. This is beyond the scope of this work and we will conduct it in future.

\section{Summary}
Energy extraction of a rotating non-Kerr black hole due to magnetic reconnection occurring in the ergosphere is investigated in this work. For a relativistic plasma rotating around a non-Kerr black hole, the magnetic field originating from the plasma could be affected by the frame dragging, resulting that anti-parallel magnetic field line could form in the ergosphere near the equatorial plane. With such anti-parallel magnetic field line configuration, a current sheet will form between the anti-parallel magnetic field and is unstable in the presence of plasmoid instability. Magnetic reconnection will happen along with available magnetic energy converted into plasma particle energy as the current sheet is destroyed by the plasmoid instability and plasmoids/flux ropes form. Magnetic reconnection produces two parts of outflows with one part accelerated in the direction of the rotation of the non-Kerr black hole and the other part accelerated in the opposite direction. Energy extraction of a rotating non-Kerr black hole will happen when the energy-at-infinity of the outflow accelerated in the opposite direction of the non-Kerr black hole is negative and the outflow accelerated in the direction of the rotation of the black hole escapes to infinity after gaining energy from the black hole.

In this work, we focus our attention on the possibility of energy extraction by magnetic reconnection in the ergosphere of a rotating non-Kerr black hole. Our results show that the deformed parameter $\epsilon$ can strongly affect the energy state of the decelerated outflow via magnetic reconnection when compared with the standard Kerr metric. When the deformed parameter is positive, the radial region in the equatorial plane in the ergosphere where energy extraction by magnetic reconnection could happen becomes larger than that for the undeformed case while the opposite effect can emerge when the deformation is negative. The radial region where energy extraction via magnetic reconnection could occur is from $r=1.2921$ to $r=1.9346$ with $a=0.9$, $\epsilon=0.3$ and $\sigma_0=10$, which is larger than that from $r=1.5927$ to $1.9138$ with $a=0.9$, $\epsilon=0$ and $\sigma_0=10$ shown in the up middle panel of Figure.~\ref{fig:figure1}. In the up right panel of Figure.~\ref{fig:figure1}, the radial region from $r=1.2883$ to $r=1.9938$ with $a=0.9$, $\epsilon=0.3$ and $\sigma_0=100$ is larger than that from $r=1.5607$ to $r=1.9925$ with $a=0.9$, $\epsilon=0$ and $\sigma_0=100$. From the middle right panel of Figure.~\ref{fig:figure1}, it shows that energy extraction could take place in the radial region from $r=1.3899$ to $r=1.9941$ with $a=0.8$, $\epsilon=0.9$ and $\sigma_0=100$, this region is larger than that from $r=1.8211$ to $1.9869$ with $a=0.8$, $\epsilon=0$ and $\sigma_0=100$. As shown in Figure.~\ref{fig:figure2}, the region of phase-space $\{r,a\}$ where energy extraction via magnetic reconnection could occur increases as the deformed parameter increases for the same plasma magnetization. Compared with the Kerr black hole, the deformed parameter with positive value can enhance the efficiency of energy extraction via magnetic reconnection and the reduced efficiency takes place when the deformed parameter is negative, the region where the efficiency of energy extraction can be enhanced shown in Figure.~\ref{fig:figure3} is the same to that shown in Figure.~\ref{fig:figure1}.

The reason why the radial region where energy extraction via magnetic reconnection with a positive deformed parameter could occur is larger than that with the undeformed case is as follows. $\epsilon^\infty_{-}$ from equation (\ref{10}) is found to be a monotonically decreasing function of $\epsilon$ when $r<2$, meaning that $\epsilon^\infty_{-}$ at a given radius (the radius is smaller than 2) decreases as $\epsilon$ increases for fixed $\sigma_0$ and $a$. With this feature of $\epsilon^\infty_{-}$ and the fixed $\sigma_0$ and $a$, $\epsilon^\infty_{-}=0$ with $\epsilon=0$ has two roots, $r_1$ and $r_2$ ($r_2>r_1$, $\epsilon^\infty_{-}(\epsilon=0,r)<0$ when $r_1<r<r_2$), $R_1$ and $R_2$ ($R_2>R_1$) are the roots of $\epsilon^\infty_{-}=0$ with a given $\epsilon>0$ and $\epsilon^\infty_{-}(\epsilon>0,r)<0$ when $R_1<r<R_2$, we find that $\epsilon^\infty_{-}(\epsilon>0,r_1)<\epsilon^\infty_{-}(\epsilon=0,r_1)=0$ and $\epsilon^\infty_{-}(\epsilon>0,r_2)<\epsilon^\infty_{-}(\epsilon=0,r_2)=0$, thus, it shows that $R_1<r_1$ and $R_2>r_2$.

Magnetic reconnection in the vicinity of supermassive black hole M87* is thought to be the mechanism powering flares observed from M87*. Since the deformed parameter could strongly affect the outflow energy via magnetic reconnection in the ergosphere of a rotating non-Kerr black hole, distinctive features of the flares may emerge as a result of the presence of hot spots produced by the merger of plasmoids forming from outflows compared with that resulting from the Kerr spacetime, providing a possible method to distinguish the rotating non-Kerr black hole from the Kerr one. Due to the fact that such features depend on general relativistic magnetohydrodynamic simulations which are beyond the scope of this work, we will perform it in future.

\section*{Acknowledgements}
I am very grateful to the anonymous referee for insightful comments that improved this work. This work is supported by the National Science Foundations of China (U1931203).



\bibliographystyle{apj}

\end{document}